# Tuning the Doping of Epitaxial Graphene on a Conventional Semiconductor *via* Substrate Surface Reconstruction

*Miriam Galbiati[1], Luca Persichetti[2,±], Paola Gori[3,+], Olivia Pulci[4,5], Marco Bianchi[6], Luciana Di Gaspare[2], Jerry Tersoff[7], Camilla Coletti[8,9], Philip Hofmann[6], Monica De Seta[2], Luca Camilli[1,4*]*

[1] Department of Physics, Technical University of Denmark, 2800 Kgs. Lyngby, Denmark

[2] Department of Sciences, Roma Tre University, 00146 Rome, Italy

[3] Department of Engineering, Roma Tre University, 00146 Rome, Italy

[4] Department of Physics, University of Rome "Tor Vergata", 00133 Rome, Italy

[5] Istituto Nazionale di Fisica Nucleare Roma 2, 00133 Rome, Italy

[6] Department of Physics and Astronomy, Aarhus University, 8000 Aarhus C, Denmark

[7] IBM Research Division, T.J. Watson Research Center, Yorktown Heights, New York, New York, 10598, USA

[8] Center for Nanotechnology Innovation @NEST, Istituto Italiano di Tecnologia, Pisa 56127, Italy

[9] Graphene Labs, Istituto Italiano di Tecnologia, Genova 16163, Italy

**Corresponding Author**

[±]: luca.persichetti@uniroma3.it; *: luca.camilli@roma2.infn.it ; [+]: paola.gori@uniroma3.it

**ABSTRACT**

Combining scanning tunneling microscopy and angle-resolved photoemission spectroscopy, we demonstrate how to tune the doping of epitaxial graphene from *p* to *n* by exploiting the structural changes that occur spontaneously on the Ge surface upon thermal annealing. Furthermore, using first principle calculations we build a model that successfully reproduces the experimental observations. Since the ability to modify graphene electronic properties is of fundamental importance when it comes to applications, our results provide an important contribution towards the integration of graphene with conventional semiconductors.

**TOC Graphic**

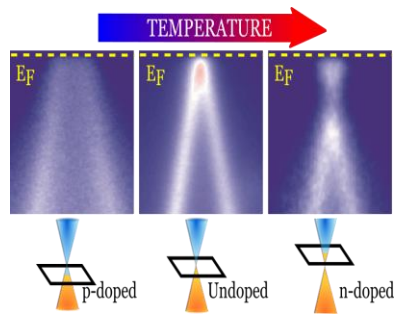

Integrating graphene into conventional semiconductor technology is expected to pave the way for the realization of novel device architectures with compelling properties [1-3]. The first experimental demonstrations of such devices include infrared image sensor arrays [4], radio frequency gas sensors [5] and vertical transistors for ultra-high frequency operations [6]. However, unlocking the true potential of these novel architectures is possible only if complete control of the full integration process is achieved. To this end, it is essential to understand the structural and electronic properties of the combined graphene/semiconductor substrate system.

Among conventional semiconductors of technological relevance, germanium (Ge) is unique as a substrate for growth of monolayer graphene [7]. It has sufficient catalytic activity with respect to the precursor gas, without a disruptive chemical affinity for carbon (in contrast to Si). Consequently the graphene/Ge system has recently attracted a great deal of interest both in materials science [8-11] and device physics [12-14]. Previous work has demonstrated that the Ge(001) surface forms high-index facets upon graphene synthesis, making the system unsuitable for further processing [12, 15-16]. In contrast, such faceting does not occur for the Ge(110) surface [17-18], which can support growth of single-crystal graphene on wafer scale [7, 19-22].

At present, the understanding of the graphene/Ge(110) interface is largely limited to its morphology, while too little is known of the electronic properties. It has been reported that samples grown *via* chemical vapor deposition (CVD) feature a hydrogen-passivated Ge surface [23] (from now on, **phase α**), and that upon annealing in vacuum above 350 °C, this surface reconstructs into a novel (6x2) phase (**phase β**) after hydrogen desorption [23-25]. It has also been shown that further in-vacuum annealing to temperatures closer to the Ge melting point leads to additional structural modifications of the Ge surface, and possibly to the formation of stronger bonds between graphene and Ge [25] (hereinafter, we refer to this phase obtained by post-growth high temperature annealing as **phase γ**). Yet, despite such morphological studies, nothing is currently known about whether and how these structural changes affect the system's electronic properties. The reported scanning tunneling spectroscopy (STS) studies have not been conclusive [26], perhaps because that technique provides very local information, and tends to be mostly sensitive to states at low $k$. Moreover, STS spectra appear to be largely dominated by features coming from the Ge substrate in this system. Thus, in order to obtain information on the band structure of the whole system on a larger scale, we use angle-resolved photoemission spectroscopy (ARPES), which has been previously used to characterize phase γ [27], as well as for a sample where the phase was not determined [28]. In addition, high temperature annealing drives in Ge the formation of defects, such as vacancies, that are known to induce significant electronic modifications in the Ge substrate [29-32]. As a result of the complex behavior and limited information, a compelling theoretical picture has not been developed yet. Earlier attempts to build models able to reproduce the experimental results required the presence of an extraordinary high dopant segregation [27].

To bridge this gap, we present here a combined scanning tunneling microscopy (STM) and ARPES study of all three graphene/Ge(110) phases mentioned above. The annealing processes induce structural changes in the interface, and we show that these in turn modify the interaction between graphene and Ge. In particular, these changes affect graphene doping, which is a crucial parameter for device applications. Furthermore, we build a model that, accounting for the presence of vacancies in Ge, successfully predicts the experimentally measured electronic properties of the system.

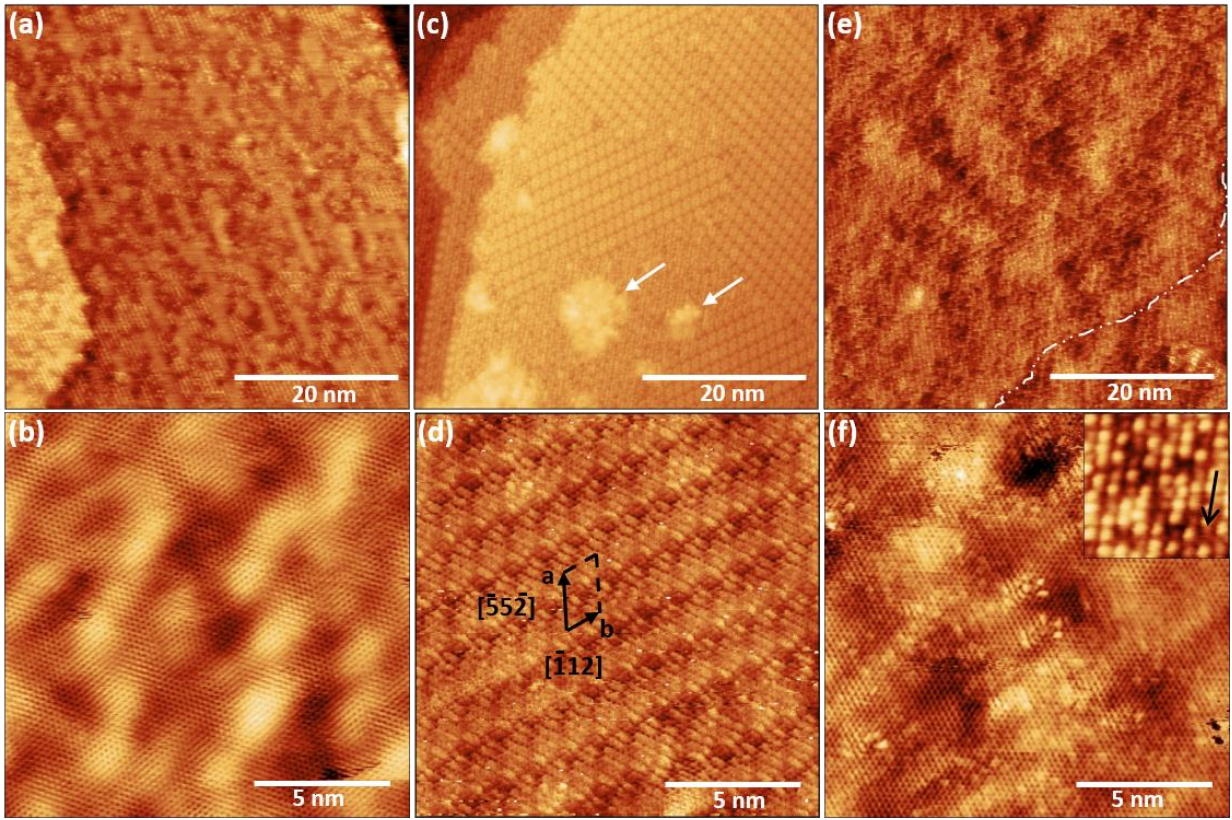

**Figure 1**. STM investigation of the graphene/Ge(110) interface for all the different phases. (a, b) STM images of phase α (V = -1.0 V and I= 0.8 nA in (a); V = 0.4 V and I = 1.0 nA in (b)). (c, d) STM images of phase β (V= -1.5 V and I= 0.8 nA in (c) and V= -1.5 V and I= 0.3 nA in (d)). (c) Coexistence of phases α and β. (d) High magnification image of phase β. In black, *a* and *b* indicate the unit cell vectors. (e, f) STM images of phase γ (V= -1.2 V and I= 0.8 nA in (e); V = -0.5 V and I = 0.8 nA in (f)). Inset of (f): Atomic-resolution STM image showing the Ge substrate. The black arrow marks the [-110] direction. Inset area is 5 x 5 nm$^2$ (V = -1.2 V and I = 0.8 nA). Figure S1 in Supporting Information shows the Fast Fourier Transform (FFT) of panel (b), (d) and (f).

Figure 1 shows STM images of the three different phases of graphene/Ge(110) interface, phase α, β and γ. When the sample is in phase α, it is possible to observe the terraces and monoatomic steps of the Ge substrate (Figure 1a). The graphene film appears to be rippled, but the graphene lattice can be clearly observed (Figure 1b). Upon annealing the sample above 350 °C in ultra-high vacuum (UHV), the H-Ge bond is broken and the Ge surface reconstructs into the phase β [24]. The size of the phase β areas depends on the duration of the annealing process. Therefore, phases α and β can coexist, as shown in the STM image reported in Figure 1c. The occasional protrusions in graphene, indicated by the arrows in Figure 1c, are nanobubbles formed by trapped hydrogen molecules that were formed upon rupture of Ge-H bond [33]. It can be noticed that they are mainly located at the Ge step edges or at the edges between areas of different phases. The atomically-resolved STM image in Figure 1d shows the unit cell of phase β and the corresponding lattice vectors (a = 2.06 nm and b = 1.30 nm). Low energy electron diffraction (LEED), Figure S2 in Supporting Information, confirms the presence of the phase β, which gives rise to a moiré pattern around graphene's primary spots. Finally, Figures 1e and 1f report characteristic STM images of phase γ, *i.e.* after annealing the sample above 700 °C in UHV conditions. This surface does not show the long-range order characteristic of phase β anymore (Figure 1e). In fact, the Ge terraces appear to be rough, with the atomic steps being hardly recognizable. (In Figure 1e, a step edge is highlighted by a dot-dashed white line just to its right). The graphene lattice is still clearly visible (Figure 1f); however, when the tunneling bias is high enough, graphene becomes transparent and the Ge lattice underneath can be imaged (Inset of Figure 1f and Figure S3 in Supporting Information). In particular, it is possible to observe that although the Ge atoms are overall aligned along the [-110] direction (marked by black arrow in the inset), many of them appear to be displaced from their lattice site either in the in-plane or out-of-plane direction. Additionally, the presence of several defects, especially vacancies, can also be noticed.

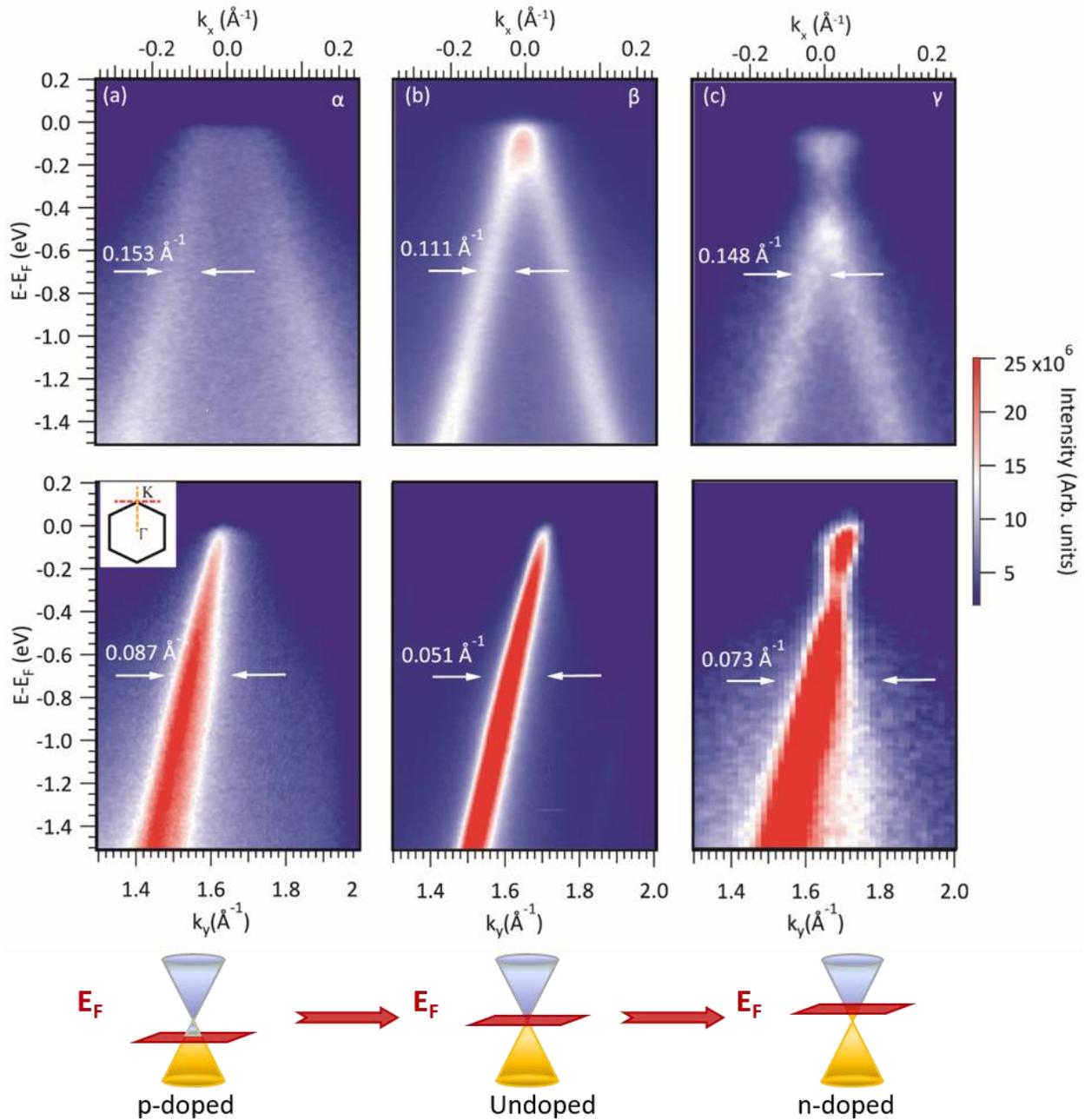

**Figure 2.** Photoemission intensity of the graphene/Ge(110) system in (a) phase α, (b) phase β, and (c) phase γ. The spectra were acquired along the direction orthogonal to the ΓK direction ((a–c) upper panels) and along the ΓK direction ((a–c) lower panels) in the Brillouin zone, as schematically shown in the inset in the middle panel in (a). Bottom row: sketch highlighting the graphene doping level in the three phases.

In order to relate the above-mentioned morphological changes to the electronic structure of the graphene/Ge interface, we performed ARPES for the three different phases. The measured dispersion, shown as photoemission intensity as a function of binding energy and *k*, is given in Figure 2a-c, with the upper (lower) panels of the figures showing data collected perpendicular (parallel) to the Γ-K direction of the graphene Brillouin zone. The crossing point of the linearly dispersing π-band branches visible in Figure 2a-c defines the position of the Dirac point ($E_D$). By linear extrapolation (see Experimental Methods in Supporting Information), $E_D$ is found at binding energy of $0.376 \pm 0.018$ eV above the Fermi energy ($E_F$) for phase α, indicating that graphene is p-doped. Further information can be extracted from the momentum distribution curve (MDC) line width, which provides insights into the graphene integrity and its interaction with the substrate. The MDC line width (FWHM) orthogonal to the Γ-K direction and taken at 0.70 eV below the Fermi level is $0.153 \pm 0.002$ Å$^{-1}$. When phase α is annealed in vacuum above 350 °C and turns into phase β, a reduction in the doping level is observed (Figure 3b). Indeed, $E_D$ is now very close to $E_F$, at a binding energy of $0.045 \pm 0.005$ eV. Thus, graphene is close to an undoped state, indicating little charge transfer with the substrate. Moreover, the MDC line width now measures 0.111 Å$^{-1}$, a value almost 30% smaller than that of phase α, indicating a weaker interaction between substrate and graphene in phase β. This finding is rather interesting, and highlights the opportunity for controlling the graphene doping and substrate interaction by suitable processing.

An additional modification of the electronic properties is found for phase γ, *i.e.* after a high temperature annealing of the sample. In this case graphene is n-doped, with $E_D$ found at $0.478 \pm 0.007$ eV below $E_F$. This experimental finding is similar to others reported ARPES measurements performed on graphene/Ge(110) after the sample was annealed at 800 °C in vacuum [27]. Previously, the best available explanation to the n-type doping was surface segregation of Sb atoms (about 1 monolayer) upon the high temperature annealing. Sb is present as dopant in the bulk of Ge also in our samples and from our STM images we note the bright patches that may appear suggestive of regions of dopant segregations. However, we rule out this interpretation as similar bright patches are seen also in the as-grown sample (compare Figure 1a and Figure 1e). For phase γ, the MDC line width is $0.148 \pm 0.002$ Å$^{-1}$, similar to that one found for phase α and larger than that of phase β. Thus, we can conclude that after the high temperature annealing, the graphene/Ge interaction is stronger than in phase β, but graphene does not degrade because a poorer structural quality would lead to a broadening in the MDC spectral line width with respect to phase α. Furthermore, we note the opening of a small band gap in graphene for phase γ [Figure 2(c) and Figure S4 in Supporting Information], induced by the stronger interaction with the substrate [25].

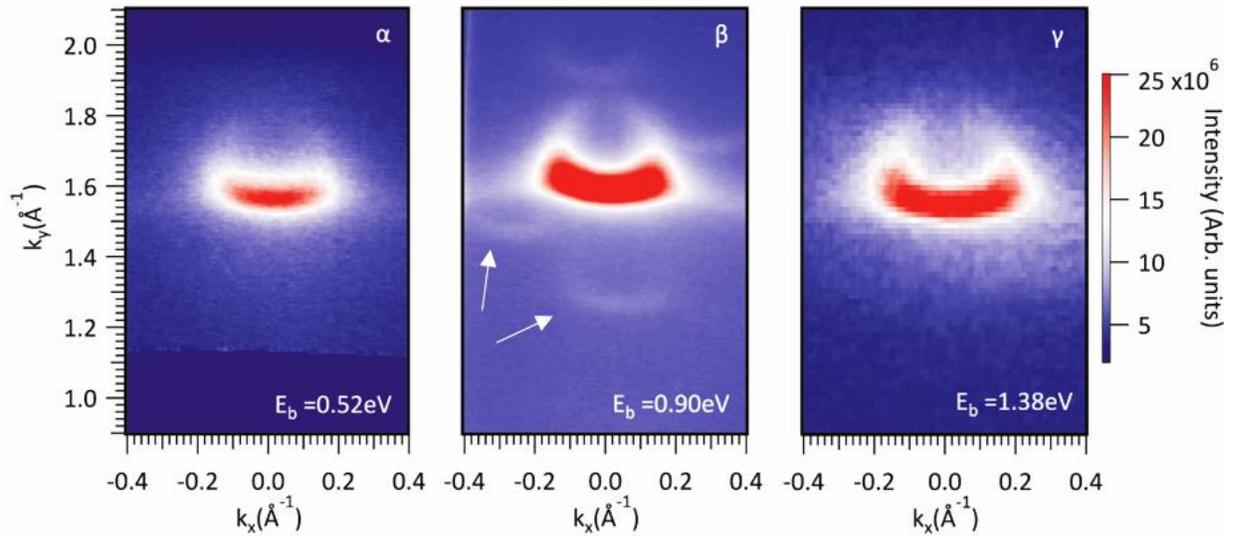

**Figure 3.** Constant binding energy cuts at 0.90 eV below the extrapolated Dirac point for phase α (left), phase β (middle), and phase γ (right). The white arrows in the middle panel point to the position of two out of four replicas.

The *k*-dependent photoemission intensity at a constant energy of 0.90 eV below $E_D$ is shown in Figure 3 for phases α (left), β (middle) and γ (right), respectively. For phase β, four replicas around the K point (we mark two of them by white arrows) can be observed along with the main Dirac cone. These replicas are due to the superperiodic potential imposed by the moiré of graphene and (6x2) Ge surface reconstruction. Accordingly, the pattern from the replicas is consistent with the LEED pattern of the corresponding sample reported in Figure S2 in Supporting Information. Interestingly, we do not observe any emerging minigaps at the crossing points of the replicas with graphene main cone, in contrast to what is found for the case of, for instance, graphene/Ir(111) [34-35]. In particular, from the analysis of the MDCs taken along $k_x$ close to the location where the replica crosses the main cone, the intensity of the spectral function does not vanish or decrease as expected in the case of a minigap (see Figure S5 in Supporting Information). Furthermore, we notice that the same constant binding energy cut shows no replicas for the other two sample phases.

To gain more insights into the system, we have built a theoretical model based on ab-initio calculations to describe phase α and phase γ (see Theoretical Methods in Supporting Information for more details), which represent the two cases showing opposite doping of graphene. We do not attempt modelling phase β because of its large cell and complexity.

Phase α is modelled by a 5-layer Ge(110) 3x5 slab, with both surfaces saturated with H, and a graphene layer on top, with a 4x8 periodicity in the rectangular supercell. The geometry is depicted in Figure 4a. Phase γ is modeled by the same supercell, but without H atoms on the top of the Ge(110) substrate (see Figure 4b). A key feature of our models is the presence of a vacancy in the surface layer of the Ge substrate (one vacancy in the 3x5 cell). The location of the vacancy is highlighted by the black circle in Figure 4a,b (additionally, the position of the missing Ge atom is also shown in Figures S6 and S7). As known in the literature [14, 29-32], when Ge is brought to high temperature (like the temperature used for graphene growth), a spontaneous formation of vacancies occurs, starting first at the surface and then spreading throughout the bulk. This confers to the Ge wafer a p-doping character regardless of the initial nominal doping. Indeed, by performing Hall measurements on different bare Ge substrates, which are nominally n-doped, we do measure p-type doping after they have been annealed in H$_2$/Ar atmosphere to the same temperature used for graphene growth (Figure S8 in Supporting Information).

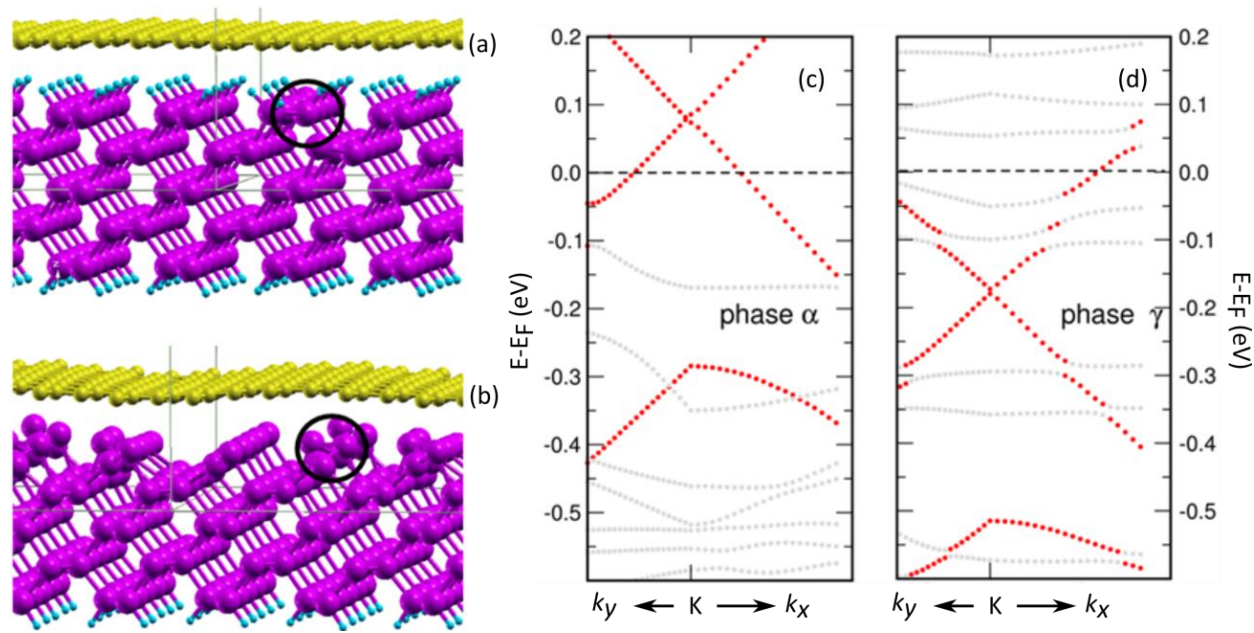

**Figure 4.** 3D side-view of phase α with all dangling bonds being H-terminated (a) and of phase γ (b). Cyan denotes H atoms, violet Ge atoms and yellow C atoms. The black circles highlight the region where the vacancy has been introduced. (c), (d): Corresponding calculated electronic band structures around K with a total $k$ range of 0.07 Å$^{-1}$. Red dots represent graphene-derived states, whereas grey dots represent Ge or H states.

The electronic band structure obtained for phase α shows that the Fermi level lies below the Dirac point, giving rise to p-doping of graphene (Figure 4c). This is qualitatively in agreement with the experiment reported in Figure 2a. To corroborate our hypothesis about the presence of vacancies in the Ge substrate, we have performed additional simulations with (i) no vacancy, and (ii) a Sb atom replacing the vacancy (Sb is the dopant nominally present in our Ge substrate). Both calculations give a Fermi level above the Dirac point, corresponding to n-doped graphene, thus at odds with experiment (see Table S1 in Supporting Information). Therefore, we believe that thermally induced vacancies and not Sb segregation play a key role in determining the doping seen in the experiment for phase α.

Phase γ was simulated by removing the H atoms on the top Ge(110) surface. In this case, we found several local minima in energy after geometrical relaxation, in contrast to the single energy minimum observed for phase α. While the most stable structure is an ordered Ge(110) surface, the other minima correspond to slightly disordered Ge surfaces (Figure 4b), with total energy within a few meV/(C atom) from the ordered one. The occurrence of a disordered surface is in fact supported by the STM images displayed in Figure 1e,f and Figure S3 in Supporting Information. The electronic band structure of this disordered phase is reported in Figure 4d, where we can see that the Fermi level lies above the Dirac point. Hence, graphene is n-doped, in agreement with the ARPES data (Figure 2c). Furthermore, we stress that a similar result – that is, n-doped graphene – is achieved when another disordered configuration of phase γ is used, where the surface Ge atoms are intentionally disordered. Instead, when an ordered Ge surface is used to simulate phase γ, no agreement with the experiment is found (see Supporting Information, Table S1). Therefore, we believe that disorder in the Ge surface and not Sb segregation plays a key role in determining the doping seen in the experiment for phase γ.

These results highlight the richness of the graphene/Ge(110) system and suggest that the experimentally observed behavior reflects not only the role of Ge-H bonds, but also of Ge vacancies and disorder, all of which evolve with increasing temperature.

In conclusion, through a combination of STM and ARPES we have experimentally demonstrated that the electronic properties of the graphene/Ge(110) system are significantly modified by temperature-driven structural changes occurring at the interface. Annealing processing can indeed be used to tune the doping of graphene *via* modification of its interaction with the Ge substrate. Notably, graphene is p-doped after CVD growth, nearly intrinsic (undoped) upon annealing above 350 °C when the Ge surface rearranges into the 6x2 reconstruction, and then n-doped if the sample is annealed above 700 °C. Starting from the STM observations we also build a theoretical model that successfully reproduces the ARPES experimental trend. Since a comprehensive understanding of the electronic properties of graphene/semiconductor interface is critical when it comes to applications, our results provide an important contribution towards the integration of graphene with conventional semiconductors.

## ACKNOWLEDGMENT

The authors are grateful to Mads Brandbyge and Fei Gao for the useful discussions. The authors thank Dr. Andrea Notargiacomo for the Hall measurements and Dr. Vaidotas Miseikis for support in growing the samples. This work is supported by the VILLUM FONDEN through the Young Investigator Program (project No. 19130) and the Centre of Excellence for Dirac Materials (Grant No. 11744). L.C. acknowledges support from the Italian Ministry of Education, University and Research (MIUR) via "Programma per Giovani Ricercatori - Rita Levi Montalcini 2017". O.P. acknowledges funding from the HORIZON2020 EU MSCA-RISE project DiSeTCom (GA 823728). The computing resources and related technical support have been provided by CRESCO/ENEAGRID High Performance Computing infrastructure (ENEA, the Italian National Agency for New Technologies, Energy and Sustainable Economic Development). C.C. acknowledges the Graphene Flagship Core 3 (contract no. 881603).

- **Supporting Information**

Supporting Information contains the Experimental and Theoretical methods; further analysis of the STM and ARPES data; LEED micrographs of phase β; Hall and resistivity measurements of Ge substrates; additional simulations.